# Developers' Leverage, Capital Market Financing, and Fire Sale Externalities: Evidence from the Thai Condominium Market

Kanis Saengchote*

*Chulalongkorn Business School*

This version: December 6, 2023

## ABSTRACT

Leveraged developers facing rollover risk are more likely to engage in fire sales. Using COVID-19 as a natural experiment, we find evidence of fire sale externalities in the Thai condominium market. Resales in properties whose developers have higher leverage ratios have lower listing prices for listed developers (who have access to capital market financing) but not unlisted developers (who primarily use bank financing). We attribute this difference to the flexibility of bank loan renegotiation versus the rigidity of debt capital market repayments and highlight the role of commercial banks in financial intermediation in the presence of information asymmetry.

Keywords: fire sale externalities, property developers, leverage, rollover risk

JEL Classification Code: G10, G18, G21, G32, R30

* Corresponding author. Chulalongkorn Business School, Chulalongkorn University, Phayathai Road, Pathumwan, Bangkok 10330, Thailand. (email: kanis@cbs.chula.ac.th).

Kanis acknowledges financial support from the Puey Ungphakorn Institute for Economic Research (PIER) Grant.

## 1. Introduction

Even in well-functioning markets, assets can be sold at deeply discounted prices relative to their fundamental value, known as a "fire sale." It is central to discussions of financial sector stability because price spillovers (negative externalities) can lead to further forced sales. After all, the concerned assets are often used as collateral.[1] For example, Acharya et al. (2007) find that recoveries of default firms tend to be lower if the industry is distressed, which the authors attribute to fire sales. This idea is tested in further detail by Benmelech and Bergman (2011), who find that bankrupt firms impose externalities by reducing the collateral value of other industry participants.

Fire sale externalities have been documented in many asset classes; for example, mutual funds (Coval and Stafford, 2007; Hau and Lai, 2017), money market funds (Schmidt, Timmermann, and Wermers, 2016), stocks (Bian et al., 2018) and bonds (Ellul, Jotikastira and Lundblad, 2011). The most influential work for real estate is Campbell, Giglio, and Pathak (2011), who find that houses sold after foreclosure are priced 27% lower on average. In a different research setting, Mian, Sufi, and Trebbi (2015) also find that foreclosures lead to a large decline in house prices and economic activities measured as residential investment and consumer demand. A more recent article by Gupta (2019) uses detailed property-level and credit bureau data to establish that foreclosure contagion can increase mortgage defaults at a very localized level.

This study aims to investigate fire sale externalities in the Thai condominium market in response to the COVID-19 shock. There are two interesting aspects of the shock. First, income shock from COVID-19 makes indebted borrowers more vulnerable and, in turn, may lead to fire sale/foreclosure externalities in the same spirit as Campbell, Giglio, and Pathak (2011), Mian, Sufi and Trebbi (2015) and Gupta (2019). Second, the panic selling in the debt market in late March 2020 led to concerns over rollover risk. Property developers have previously relied on bank-financed construction loans for developments. However, since 2013, the issuance of bonds and commercial papers by property development companies has become more popular.

To see why this is potentially alarming, we discuss the characteristics of construction loans in Thailand compared to fixed-income securities when used to finance for sale developments. Construction loans are like credit lines with pre-specified conditions for drawdowns and repayments. Lenders typically require developers to co-invest with a loan-to-cost ratio of around 50-60% and may ask developers to achieve minimum presale requirements before approval. These mechanisms intend to minimize adverse selection and moral hazard problems in construction management and are often used in project financing. For drawdowns, verification of construction progress is necessary for disbursements, which helps reduce misuse of capital and moral hazard. During the construction stage, interests are accrued, and no repayment is necessary.

Once completed, developers are incentivized to repay as soon as possible to reduce interest costs. They are also contractually required to repay as a percentage of sales proceeds, often called a "sweep," which helps reduce credit risk. In sum, bank financing is designed to address information asymmetry inherent in financial intermediation and performs a necessary role in costly state verification (see, for example, Townsend, 1979; Campbell and Kracaw, 1980; Diamond, 1984; Fama, 1985).

---

[1] Popular models of fire sale externalities are Shleifer and Vishny (1992) and Kiyotaki and Moore (1997). For a review of fire sales in finance and macroeconomics, see Shleifer and Vishny (2011).

On the other hand, debt capital market financing (e.g., bonds and commercial papers) tends to be less restrictive and does not formally verify the use of proceeds. For many developers, capital market financing is more attractive than construction loans as there are fewer restrictions, and borrowing rates tend to be lower, particularly for securities that receive "investment grade" designation from rating agencies. However, the lump sum disbursement is prone to adverse selection and moral hazard, while periodic coupon payments impose cash flow burdens with no inflows during development stages. Consequently, only developers with an established track record and ongoing cash flows have access to this type of financing. In addition, developers are often unable to match cash flow duration and thus are exposed to rollover risk when they take on short-maturity debt to save interest costs. Rollover risk may be less of a concern in normal times, but certain circumstances can expose the developers to severe shortfalls.

At the end of 2019, listed developers had THB 611 million in interest-bearing debt outstanding, and THB 375 million was capital market financing, representing 61.4% of total debt. About a third of that amount was due in 2020. According to the Thai Bond Market Association, as of April 26, 2020, THB 148 billion commercial papers and bonds were due in the next 12 months, representing more than 16% of all fixed-income securities due by April 2021. This rollover risk leads to some developers resorting to deep discounts on their inventories (primary market) to repay their bonds, which can spill over to the secondary market. In other words, developers' financial positions may lead to fire sale externalities in the housing market.

In this paper, we use more than 350,000 secondary market condominium listings of more than 1,000 developments by 184 developers in Bangkok between January 2019 and September 2020 to investigate how developers' leverage and rollover risk in capital market financing affects secondary condominium market, using COVID-19 as a natural experiment. While we do not directly observe developers' actions, resale activities are useful to infer fire developers' fire sales from the price externalities. We find evidence consistent with fire sale externalities among listed developers with high leverage but not unlisted developers. We attribute this difference to the flexibility of bank loan renegotiation versus the rigidity of debt capital market repayments. We also demonstrate that developers' inventory (*excess inventory channel*) and the need for capital market refinancing (*rollover risk channel*) affect fire sales externalities.

To our knowledge, our study is the first to investigate how corporate financial policies affecting strategic choices in the primary market can unintentionally lead to externalities in the secondary market. Our contribution is to demonstrate that policymakers can benefit from monitoring developers' leverage and their debt compositions. With high-profile examples in China, such as China Evergrande, which defaulted on its bonds in December 2021,[2] and the overall rollover risk continuing into 2023,[3] our study highlights the inherent risk in debt capital market financing that can build up during periods of relaxed monetary policy.

---
[2] https://www.nytimes.com/2021/12/09/business/china-evergrande-default.html

The rest of the paper is organized as follows: Section 2 provides an overview of the literature on fire sales and develops the hypotheses for this study. Section 3 describes data sources and the empirical methodology. Section 4 presents and discusses the results, concluding in Section 5.

## 2. Hypothesis Development

Fire sale externalities and their impact on the real economy have long been discussed in the macroeconomic literature since Fisher (1933). However, the theoretical foundations behind more recent discussions draw their insights from the microeconomic model of Shleifer and Vishny (1992), where debtors are forced to sell assets at "dislocated" prices in order to repay debt and the macroeconomic model of Bernanke and Gertler (1989), Kiyotaki and Moore (1997), and Bernanke, Gertler, and Gilchrist (1999) where credit cycles are driven by self-reinforcing changes in asset values and availability of collateralized credit.

Empirical studies of fire sale externalities tend to focus on specific asset classes. For example, Coval and Stafford (2007) and Hau and Lai (2017) find that redemption shocks can lead to forced selling of equity mutual funds' holdings. Even cash-like investments, such as money market funds, are not free from such redemption shock, as Schmidt, Timmermann, and Wermers (2016) show. Ellul, Jotikastira, and Lundblad (2011) investigate insurance companies (whose investment opportunities are strictly regulated) and find that they are forced to sell downgraded corporate bonds at discounted prices. These sudden demand shocks tend to occur with the lack of counterparties, leading creditors to anticipate this in their credit pricing, as shown in the U.S. airline industry by Benmelech and Bergman (2009). Fire sales can also propagate across banks' balance sheets, as demonstrated in a theoretical model and a cross-sectional empirical investigation by Greenwood, Landier, and Thesmar (2015). Duarte and Eisenbach (2019) extend this insight and develop an index of aggregate vulnerability to fire sales (AV index), which they show can be a useful five-year-ahead early warning indicator.

This paper focuses on the housing market, intricately linked to the real sector and financial markets. The most influential article on fire sale externalities in the housing market is by Campbell, Giglio, and Pathak (2011), who find that houses sold after foreclosure are priced 27% lower on average, and the discount is related to distance from distressed properties. Foreclosure externalities is a well-researched topic that has gained popularity since the 2008-2009 crisis because of its economic ramifications. See, for example, Shuetz, Been, and Allen (2008), Harding, Rosenblatt, and Yao (2009), Lin, Rosenblatt and Yao (2009), Rogers and Winter (2009), Anenberg and Kung (2014), Gerardi et al. (2015) and Gupta (2019). However, these articles tend to focus on external shocks at the macroeconomic level (e.g., general economic condition, lending crunch) or idiosyncratic (e.g., death, individual bankruptcy) and how they affect transaction prices without the role and actions of developers. Given the durability of real estate assets and substitutability between new and pre-owned assets, we

---

[3] https://asia.nikkei.com/Spotlight/Caixin/Chinese-developers-facing-141-billion-in-maturing-bonds-this-year

introduce the interplay between the primary and secondary market and how it may exacerbate fire sales.[4]

We hypothesize that heavily leveraged developers are more inclined to sell their existing inventory at discounted prices, which creates additional price pressure on secondary market listings. In addition, many listed developers in Thailand rely on capital market financing whose repayment terms are not tied to cash flows from sales like construction loans. Consequently, developers who face rollover risk may be more under pressure, thus exerting pressure on the secondary market.

### 3. Data and Methodology

#### 3.1. Sample

We obtained property listing data (resale of completed units only) in Bangkok and surrounding areas between January 2019 and June 2021 from Baania.com, a Thai real estate data platform. Because users voluntarily provide unit-level data, data consistency varies, and much of the listing details are provided as unstructured text, so we limit our analysis to common variables such as unit size and number of bedrooms and fill missing unit size with median value for the bedroom category where applicable.[5] For development-level data such as completion date, number of units, number of floors, and geocoded locations, the platform provides manually verified information for popular developments, and we manually fill out missing data with Internet searches. The geographical distribution of listing prices is visualized in Figure 1, and the histogram of pre-COVID listing prices is in Figure 2.

We supplement listing data by developers' financial positions obtained from the Stock Exchange of Thailand for listed developers and hand-collected from the Department of Business Development, Ministry of Commerce for unlisted developers. We require all developers to have information on sales, cost of goods sold, total assets, total liabilities, and inventory to calculate the leverage (defined as total liabilities divided by total assets) and inventory ratios (compared to cost of goods sold and measured in years). Many listed developers have access to capital market financing and thus have issued short-term commercial papers (270 days or less in Thailand) and longer-term bonds in the run-up to COVID-19. We obtain fixed-income securities from the Thai Bond Market Association (ThaiBMA), a self-regulatory organization that also functions as an information center for the bond market.

Our final sample includes 355,032 listings from 1,075 developments of 184 developers, 54 of whom are listed in the stock market. 68.1% of the developments in the sample are from listed companies, with the top three developers accounting for 30%. The summary statistics of the units, developments, and developers are reported in Table 1. On average, listed developers are about ten times larger than unlisted developers, have similar leverage ratios, and have half as much inventory.

---

[4] A similar argument regarding the relationship between primary market and secondary market for durable assets is made in Noparumpa and Saengchote (2017).

[5] Many studies that use hedonic pricing regression use property attributes that are verified by assessors and maintained in government's databases for administrative purposes (e.g. property tax assessment). Such information is not digitized and readily available in Thailand, so we base our analyses on variables that can be consistently and reliably obtained. The lack of control variables limits the ability of hedonic pricing regressions, but because our analyses are based on condominiums rather than houses, there is less heterogeneity, alleviating potential concerns raised by this issue. To further reduce concerns, we restrict our analyses to 1- and 2-bedroom units, which tend to be much more homogenous within a development. 1-bedroom units account for 70% of the data, while 2-bedroom 26%, so by dropping larger units, we only lose 4% of the sample.

**Figure 1: Geographical distribution of property prices in Bangkok**

This figure plots the locations of condominium developments in Bangkok and surrounding areas in our sample. The average price per square meter of units listed for sale between 2019Q2 and 2020Q1 is calculated and rounded to the nearest THB 1,000 per square meter. The color of each point depicts the price range divided into six categories, with blue representing the highest average price and red the lowest. Each hexagonal block represents 0.2 square kilometers, and block height represents the number of condominium developments in that block. The median price per square meter in each block is computed and divided into six ranges, with blue also representing the highest median price and red the lowest.

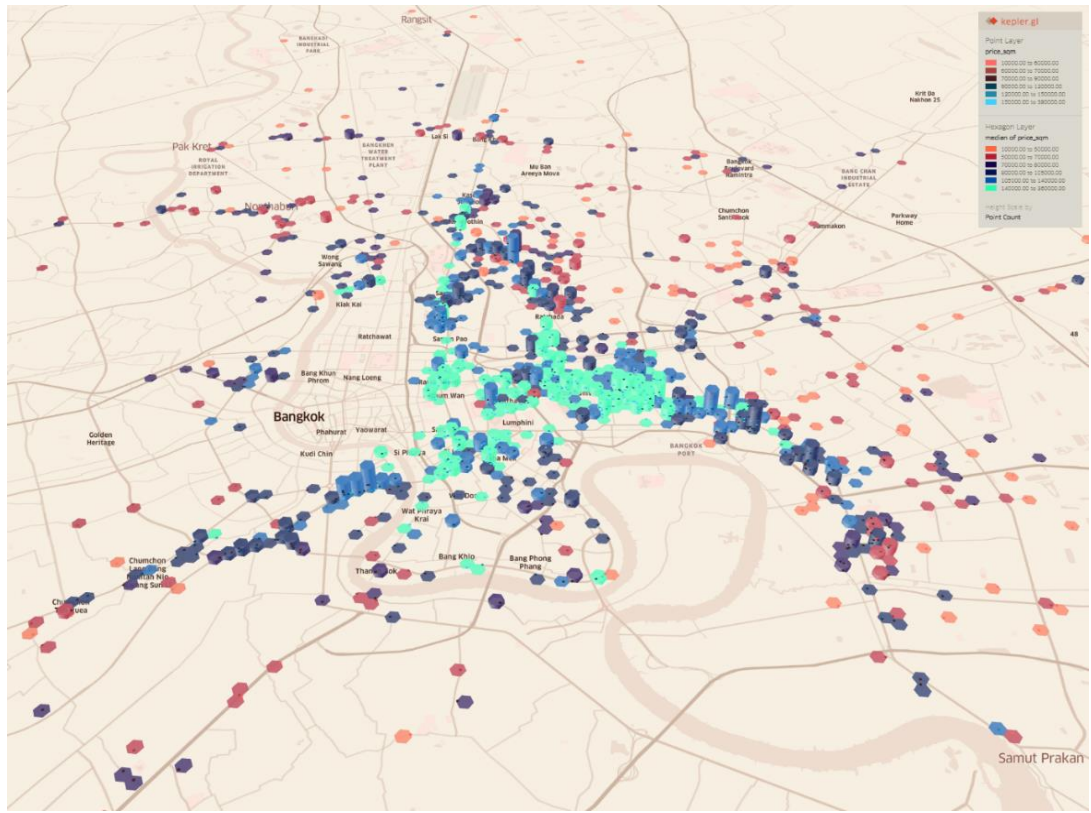

**Figure 2: Histogram of property prices in Bangkok**

This figure plots the distribution of average price per square meter of units listed for sale between 2019Q2 and 2020Q1 (pre-COVID), using the same data as Figure 1.

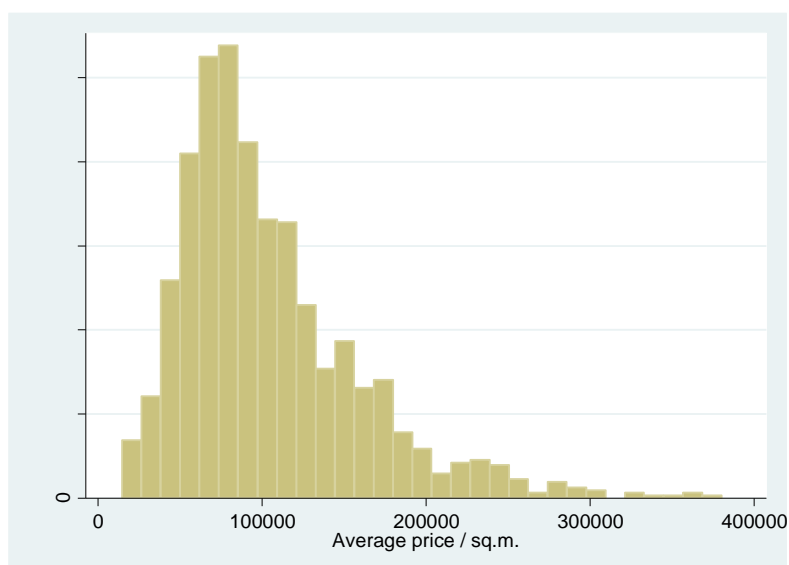

**Table 1: Summary statistics**

Panel A reports the summary statistics of listings between January 2019 and June 2021. Listing data is obtained from Baania.com. Missing unit size is replaced by the median value for the units in the same development with the same number of bedrooms. Summary statistics for listing data are divided into three periods: pre-closure (January 2019 to March 2020), closure to Q3 (April 2020 to September 2020), and post-Q3 (October 2020 to June 2021). Panel B reports the summary statistics for the condominium developments, and panel C reports developers' financial information as of 2019. Commercial papers and bonds due (collectively called "debt") are calculated with the end of March 2020 as the base date to coincide with the closure order and the bond mutual fund run. For the proportion of debt due, only listed developers are included. 37 out of 54 listed developers use capital market financing.

Panel A: listing data

|  |  | Price/unit | Price/sq.m. | Unit size (sq.m.) | % Two bedrooms |
|---|---|---|---|---|---|
| Pre-closure | Mean | 6,483,390 | 136,795 | 43.59 | 24.4% |
| Num observations | Std Dev | 5,977,654 | 68,831 | 20.30 | |
| 131,302 | Median | 4,780,000 | 123,282 | 35.00 | |
| | | | | | |
| Closure - Q3 | Mean | 4,819,355 | 109,870 | 39.85 | 19.2% |
| Num observations | Std Dev | 5,401,797 | 62,604 | 17.75 | |
| 21,245 | Median | 3,220,000 | 91,667 | 33.67 | |
| | | | | | |
| Post Q3 | Mean | 6,636,202 | 134,935 | 45.07 | 27.5% |
| Num observations | Std Dev | 6,293,842 | 72,196 | 21.42 | |
| 202,485 | Median | 4,850,000 | 120,000 | 37.00 | |
| | | | | | |
| All periods | Mean | 6,470,968 | 134,123 | 44.21 | 25.9% |
| Num observations | Std Dev | 6,142,624 | 70,693 | 20.85 | |
| 355,032 | Median | 4,700,000 | 119,490 | 35.50 | |

Panel B: development data

|  | Mean | Std Dev | Median | Count |
|---|---|---|---|---|
| Year finished | 2014.1 | 4.81 | 2015 | 976 |

| Missing year finished | 9.2% | | | 1,075 |
|---|---|---|---|---|
| Num units | 559 | 590 | 402 | 1,075 |
| Num floors | 21 | 13.4 | 20 | 1,075 |
| Developer listed in SET | 68.1% | | | 1,075 |
| Population density (Facebook) | 2014.1 | 4.81 | 2015 | 976 |

Panel C: developer data

| Unlisted Developers (N=130) | Mean | Std Dev | Median |
|---|---|---|---|
| Total sales (THB million) | 1,142.3 | 4,668.2 | 97.4 |
| Total assets (THB million) | 3,330.3 | 10,722.7 | 514.1 |
| Total liabilities (THB million) | 2,048.0 | 6,100.1 | 280.0 |
| Inventory (THB million) | 1,795.0 | 8,343.8 | 218.5 |
| Leverage ratio (TL/TA) | 58.8% | 35.9% | 60.1% |
| Inventory (years) | 10.50 | 14.44 | 4.19 |

| Listed Developers (N=54) | Mean | Std Dev | Median |
|---|---|---|---|
| Total sales (THB million) | 9,052.8 | 10,189.8 | 4,018.6 |
| Total assets (THB million) | 32,873.1 | 34,949.3 | 15,993.1 |
| Total liabilities (THB million) | 19,781.7 | 22,782.8 | 9,623.3 |
| Inventory (THB million) | 14,714.7 | 17,337.1 | 8,202.2 |
| Leverage ratio (TL/TA) | 57.2% | 17.3% | 59.2% |
| Inventory (years) | 4.78 | 7.29 | 3.22 |
| Bond issuer | 68.5% | | |
| Debt due in 6 months / TL | 7.1% | 8.8% | 2.9% |
| Debt due in 12 months / TL | 13.1% | 15.0% | 8.4% |

### 3.2. Empirical Strategy

Our empirical strategy is to combine the hedonic pricing regression (Rosen, 1974) with the difference-in-differences (DiD) strategy in repeated cross-section, similar to Qian et al. (2021), who study the impact of COVID-19 on housing prices in China. Because Thailand issued a closure order in early April 2020, we defined April 2020 and the period after that as the post-period in the DiD framework. In March 2020, there was also a bond mutual fund run. As a result, the credit spread spiked, leading to concerns about rollover risk.[6] Developers with scheduled repayments would likely need to sell at deep discounts to meet debt obligations, competing against secondary market resales. It is important to emphasize that we do not observe primary market activities, so we use resale activities to infer their actions. The DiD treatment is the extent of leverage that would compel developers to sell their inventories as external financing dries up.

$$y_{idt} = \alpha_d + \delta_t + \beta_1 Post_t \cdot Leverage_i + \gamma' X_i + \varepsilon_{idt} \qquad (1)$$

---

[6] https://www.thaibma.or.th/Doc/annual/SummaryMarket2020.pdf

Our regression equation follows Equation 1, where the dependent variable $y_{ist}$ is the log listing price of unit $i$ in development $d$ at time $t$. Because of limited data, we use fixed effects $\alpha_d$ for each development to control for unobservable development-specific variations (e.g., brand, neighborhood) that could influence prices, and $\delta_t$ for each listing month that controls for general movements in property prices. The inclusion of both fixed effects means the treatment effect is identified by the coefficient on the DiD interaction of $Post_t$ and $Leverage_i$.[7] The control variables $X_i$ unit attributes comprise unit size in sq.m., an indicator variable for 2-bedroom units, and indicator variables for property age at the time of listing. Development- and developer-specific characteristics are subsumed by the fixed effects.

We extend the DiD framework to investigate various other channels. For example, indebted developers with more inventories may be more willing to sell at lower prices, as the remaining inventory has real option value (*excess inventory channel*). Consequently, the resale properties of such developers may face greater pressure. We compute inventory ratios (measured in years) by dividing outstanding inventory by the cost of goods sold using 2019 data.

For listed companies, we investigate whether the need for capital market refinancing affects their behavior (*rollover risk channel*). Of the 54 listed developers, 37 issue bonds and commercial papers ('debt'). We compute the ratio of debt due in 6 months and 12 months from the end of March 2020 to total liabilities.

The fire sale externalities predict inventory and debt-due ratios to be negatively related to the prices of secondary market listings. In all analyses, standard errors are clustered at the development level.

## 4. Results

### 4.1. Leveraged Fire Sale Externalities

We begin by discussing the summary statistics presented in Table 1 Panel A. The average listing prices post-closure compared to the subsequent period declined from THB 6.48 million to THB 4.82 million. Listed units are smaller (39.9 square meters versus 43.6 square meters pre-closure) and are more likely to be 1-bedroom listings (19.2% compared to 24.4%). The average price per square meter also declined from THB 136,795 to THB 109,870, so the government closure order decreased condominium prices. However, the objective of our study is not to document a price decline but to investigate whether prices in condominiums of developers who face financial pressure decline more.

**Table 2: Developers' Leverage and Fire Sale Externalities**

This table reports the result from the difference-in-differences regressions of log listing prices between January 2019 and September 2020. Because of limited data availability, control variables only include unit size (measured in square meters), an indicator variable for listings with two bedrooms (only 1- and 2-bedroom units are included), and indicator variables for development age but are omitted for brevity. Post is an indicator variable for listings from April 2020 to September 2020. For interaction terms, leverage is defined as total liabilities divided by total assets. Column 1 includes properties by both unlisted and listed developers. Columns 2 and 3 divide the sample into properties by unlisted and listed developers, respectively. All regressions include development and month fixed effects. Standard errors are clustered at the development level. Stars correspond to the statistical significance level, with *, **, and *** representing 10%, 5%, and 1%, respectively.

---

[7] We measure leverage as total liabilities divided by total assets, as the financial statements of unlisted companies filed to the Department of Business Development are less detailed compared to listed counterparts, so conventional ratios based on interest-bearing debt are not possible.

| Sample | (1) All | (2) Unlisted | (3) Listed |
|---|---|---|---|
| Post * Leverage | -0.0546*** | -0.0169 | -0.0950*** |
|  | (0.017) | (0.024) | (0.025) |
| Unit size (sq.m.) | 0.0157*** | 0.0155*** | 0.0158*** |
|  | (0.000) | (0.001) | (0.000) |
| Two bedrooms | 0.1207*** | 0.1017*** | 0.1244*** |
|  | (0.010) | (0.020) | (0.011) |
| Observations | 152,547 | 29,990 | 122,557 |
| Adj R-squared | 0.957 | 0.952 | 0.957 |

We turn to the DiD analysis for the periods immediately following the closure order, reported in Table 2. The baseline results of the DID analyses are reported in Columns 1 to 3, where the first column pools units from all developers, the second column includes only units from unlisted developers, and the third column includes only units from listed developers. Both $Post_t$ and $Leverage_i$ are subsumed by the time fixed effects and development fixed effects, so only the interacted DiD variable is identified. Including various fixed effects and control variables results in high adjusted R-squared values, reflecting a stringent identification strategy.

The interacted DiD coefficient is negative as predicted and statistically significant at the 1% level. On average, the resale unit of developers with a leverage ratio of one standard deviation higher would list for 1.7% less. The leveraged fire sale effect is only present in properties by listed developers, and the coefficient of -0.095 translates into a 1.6% price impact per one stand deviation change in leverage ratio. We attribute this difference to the methods of debt financing for listed and unlisted developers.

While unlisted developers can issue bonds and commercial papers in principle, most of the issuers in Thailand are listed companies. Bank lending is more easily negotiable than capital market financing, and this flexibility is more valuable during distress. Piskorski, Seru, and Vig (2010) demonstrate that bank-held loans are less likely to be foreclosed than securitized loans because of renegotiation friction. As discussed earlier, 61.4% of listed developers' interest-bearing debt comprises bonds and commercial papers with a diffuse investor base. The rigidity of debt capital market repayments likely spurs fire sales.[8]

### 4.2. Excess Inventory Channel

Next, we investigate the influence of unsold inventories on fire sale externalities. Developers with unsold inventory may be more willing to offload their inventory, as the remaining inventory still has real option value. The coefficients are only statistically significant for listed developers. Double-interaction models are not simple to interpret, and the treatment effect depends on the levels of variables. The marginal effect of increasing the inventory ratio can be computed from the coefficients of Post * Inventory (years) and Post * Leverage * Inventory (years), which can be expressed as 0.0037 – 0.0065 * Leverage. The contribution becomes negative as the leverage ratio exceeds 56.9% (just under the median).

---
[8] Bank loan renegotiations can also suffer from holdouts, as Brunner and Krahnen (2008) find that bank loan restructuring is more difficult when there are multiple lenders. However, residential real estate developments in Thailand are typically financed by a few lenders. Moreover, the Thai banking system consists of a small number of banks, making renegotiation easier when needed.

The *excess inventory channel* is only relevant for developers with high leverage. To see the economic impact, consider the leverage ratio for the 75th percentile developer of 70.6%. For this developer, increasing the inventory ratio from the average value of 4.78 years by one standard deviation will reduce the listing prices of the developer's resale units by 1.1%. Like the result in Table 2, unlisted developers' resale units are unaffected by leveraged fire sales, corroborating the view of greater flexibility in bank loan renegotiation.

**Table 3: Fire Sale Externalities and Inventory**

This table reports the result from the difference-in-differences regressions of log listing prices between January 2019 and September 2020. Because of limited data availability, control variables only include unit size (measured in square meters), an indicator variable for listings with two bedrooms (only 1- and 2-bedroom units are included), and indicator variables for development age but are omitted for brevity. Post is an indicator variable for listings from April 2020 to September 2020. For interaction terms, leverage is defined as total liabilities divided by total assets, and inventory (in years) is defined as inventory in December 2019 divided by the cost of goods sold in 2019. Column 1 includes properties by both unlisted and listed developers. Columns 1 and 2 divide the sample into properties by unlisted and listed developers, respectively. All regressions include development and month fixed effects. Standard errors are clustered at the development level. Stars correspond to the statistical significance level, with *, **, and *** representing 10%, 5%, and 1%, respectively.

|  | (1) | (2) |
|---|---|---|
| Sample | Unlisted | Listed |
| Post * Leverage | -0.0043 | -0.0657** |
|  | (0.025) | (0.029) |
| Post * Inventory (years) | 0.0007 | 0.0037 |
|  | (0.001) | (0.002) |
| Post * Leverage * Inventory (years) | -0.0014 | -0.0065** |
|  | (0.002) | (0.003) |
| Unit size (sq.m.) | 0.0155*** | 0.0158*** |
|  | (0.001) | (0.000) |
| Two bedrooms | 0.1016*** | 0.1245*** |
|  | (0.020) | (0.011) |
| Observations | 29,990 | 122,557 |
| Adj R-squared | 0.952 | 0.957 |

### 4.3. Rollover Risk Channel

This section delves deeper by distinguishing listed developers with immediate refinancing needs. We interact the DiD variables with the ratio of capital market debt due in 6 and 12 months and preserve the interaction with the inventory ratio from Section 4.2. The results are reported in Table 4. The coefficients for both 6 and 12 months are negative and statistically significant at the 5% level. Like earlier, we identify the leverage thresholds where the marginal effect of the second interaction term is negative. The threshold is 65.1% for debt due in 6 months and 60.2% for debt due in 12 months.

Thus, the *rollover risk channel* is also only relevant for developers with high leverage. Using the leverage ratio at the 75th percentiles as before, increasing debt due in 6 months from the average by one standard deviation will reduce listing prices by 0.7%, and for debt due in 12 months, it is 1.1%. Our result demonstrates that refinancing needs during funding draught can exacerbate fire sale externalities.

**Table 4: Fire Sale Externalities and Capital Market Financing**

This table reports the result from the difference-in-differences regressions of log listing prices between January 2019 and September 2020. Because of limited data availability, control variables only include unit size (measured in square meters), an indicator variable for listings with two bedrooms (only 1- and 2-bedroom units are included), and indicator variables for development age but are omitted for brevity. Post is an indicator variable for listings from April 2020 to September 2020. For interaction terms, leverage is defined as total liabilities divided by total assets. Debt due is calculated as the face value of commercial papers and bonds due in the next 6 months (Column 1) or 12 months (Column 2), with March 2020 as the baseline date. All regressions include development and month fixed effects. Standard errors are clustered at the development level. Stars correspond to the statistical significance level, with *, **, and *** representing 10%, 5%, and 1%, respectively.

|  | (1) 6 months | (2) 12 months |
|---|---|---|
| Post * Debt due in … months | 0.5505** | 0.2295* |
|  | (0.249) | (0.118) |
| Post * Leverage * Debt due in … months | -0.8451** | -0.3811** |
|  | (0.393) | (0.193) |
| Post * Leverage | 0.0795 | 0.0287 |
|  | (0.076) | (0.062) |
| Post * Inventory (years) | 0.0073** | 0.0040 |
|  | (0.003) | (0.003) |
| Post * Leverage * Inventory (years) | -0.0113*** | -0.0069** |
|  | (0.004) | (0.004) |
| Unit size (sq.m.) | 0.0158*** | 0.0158*** |
|  | (0.000) | (0.000) |
| Two bedrooms | 0.1246*** | 0.1246*** |
|  | (0.011) | (0.011) |
| Observations | 122,557 | 122,557 |
| Adj R-squared | 0.957 | 0.957 |

### 4.4. Post-Intervention Reversal

So far, we have only analyzed listing prices up to the end of 2020Q3. In this section, we extend the analysis by one year to 2021Q2. There were 202,485 listings during that period, more than the entire period of the earlier sample, suggesting that the real estate market has bounced back. Average listing prices (per unit and per sq. m.) also returned to pre-closure levels. We use the double-interaction models from Section 4.2 and Section 4.3, creating another indicator for listings after 2020Q3 until 2022Q2 to assess the longer-term effect and report the results in Table 5.

For the post-closure period (Up to Q3), the results remain largely the same, with leverage only affecting listed companies (Column 2), and the leverage thresholds for the excess inventory channel and rollover risk channel around 65% for 6 months and 60% for 12 months as before. For the subsequent period (After Q3), the double-interaction coefficients for listed companies (Columns 3 and 4) become statistically insignificant, and the leverage thresholds for the rollover risk channel increase to more than 80%. The absence of the leveraged fire sale externalities suggests that the rollover risk concern was allayed.

**Table 5: Post-Intervention Reversal**

This table reports the result from the difference-in-differences regressions of log listing prices between January 2019 and June 2021. Because of limited data availability, control variables only include unit size (measured in square meters), an indicator variable for listings with two bedrooms (only 1- and 2-bedroom units are included), and indicator variables for development age but are omitted for brevity. Up to Q3 is an indicator variable for listings from April 2020 to September 2020, and After Q3 is an indicator variable for listings from October 2020 to June 2021. Columns 1 and 2 divide the sample into properties by unlisted and listed developers, respectively. For interaction terms, leverage is defined as total liabilities divided by total assets. Debt due is calculated as the face value of commercial papers and bonds due in the next 6 months (Column 3) or 12 months (Column 4), with March 2020 as the baseline date. All regressions include development and month fixed effects. Standard errors are clustered at the development level. Stars correspond to the statistical significance level, with *, **, and *** representing 10%, 5%, and 1%, respectively.

| Sample | (1) Unlisted | (2) Listed | (3) 6 months | (4) 12 months |
|---|---|---|---|---|
| Up to Q3 * Leverage | 0.0067 | -0.0538* | 0.0742 | 0.0271 |
|  | (0.024) | (0.028) | (0.070) | (0.059) |
| Up to Q3 * Inventory (years) | 0.0003 | 0.0036 | 0.0071** | 0.0042 |
|  | (0.001) | (0.002) | (0.003) | (0.003) |
| Up to Q3 * Leverage * Inventory (years) | -0.0008 | -0.0063** | -0.0111*** | -0.0071** |
|  | (0.002) | (0.003) | (0.004) | (0.003) |
| Up to Q3 * Debt due in … |  |  | 0.4785** | 0.1998* |
|  |  |  | (0.232) | (0.113) |
| Up to Q3 * Leverage * Debt due in … |  |  | -0.7152* | -0.3356* |
|  |  |  | (0.370) | (0.187) |
| After Q3 * Leverage | 0.0209 | -0.0377* | 0.0654 | 0.0406 |
|  | (0.015) | (0.020) | (0.040) | (0.037) |
| After Q3 * Inventory (years) | 0.0012* | -0.0015 | 0.0023 | 0.0003 |
|  | (0.001) | (0.002) | (0.002) | (0.002) |
| After Q3 * Leverage * Inventory (years) | -0.0017** | 0.0014 | -0.0036 | -0.0010 |
|  | (0.001) | (0.003) | (0.003) | (0.003) |
| After Q3 * Debt due in … |  |  | 0.3323** | 0.1491** |
|  |  |  | (0.144) | (0.075) |
| After Q3 * Leverage * Debt due in … |  |  | -0.4018 | -0.1861 |
|  |  |  | (0.245) | (0.128) |
| Unit size and 2-bedroom dummy | Yes | Yes | Yes | Yes |
| Observations | 68,913 | 286,119 | 286,119 | 286,119 |
| Adj R-squared | 0.953 | 0.959 | 0.959 | 0.959 |

The bond mutual fund run in March 2020 spurred immediate actions. In a few days after the run, the Bank of Thailand, the Securities and Exchange Commission, and the Ministry of Finance held a joint press conference, announcing measures to facilitate new corporate issuances and mitigate rollover risk, such as the Corporate Bond Stabilization Fund (BSF) that will be supported by Thai commercial banks, insurance providers, and the Government Pension Fund. Bond credit spreads widened after the run and began tilting downward toward the end of the year.[9] The ThaiBMA 2020 Annual Report states, "Although the [BSF] facility has not been utilized so far, it has helped restore investor confidence during the time of volatility." The swift response successfully alleviated the concerns as intended.

As capital market financing resumed, the initial advantage of unlisted developers with bank financing began to wane. Column 2 of Table 5 shows that unlisted developers previously unaffected by leverage experienced fire sale externalities in the preceding year. The coefficients imply a leverage threshold of 70.6%, quite high compared to the average for unlisted developers, but it points toward a reversal of fortunes.

**Figure 3: Bonds and Commercial Papers Issuance by Developers**

This figure plots the quarterly sum of bond and commercial papers issued by property developers between 2018Q1 and 2023Q3. Data is obtained from the Thai Bond Market Association (ThaiBMA), a self-regulatory organization that also functions as an information center for the bond market. Debt maturing in one year or less is typically commercial papers, a zero-coupon instrument with maturity of 270 days or less in Thailand. Maturity is divided into four categories: due within one year, due between one to three years, due between three to five years, and due more than five years.

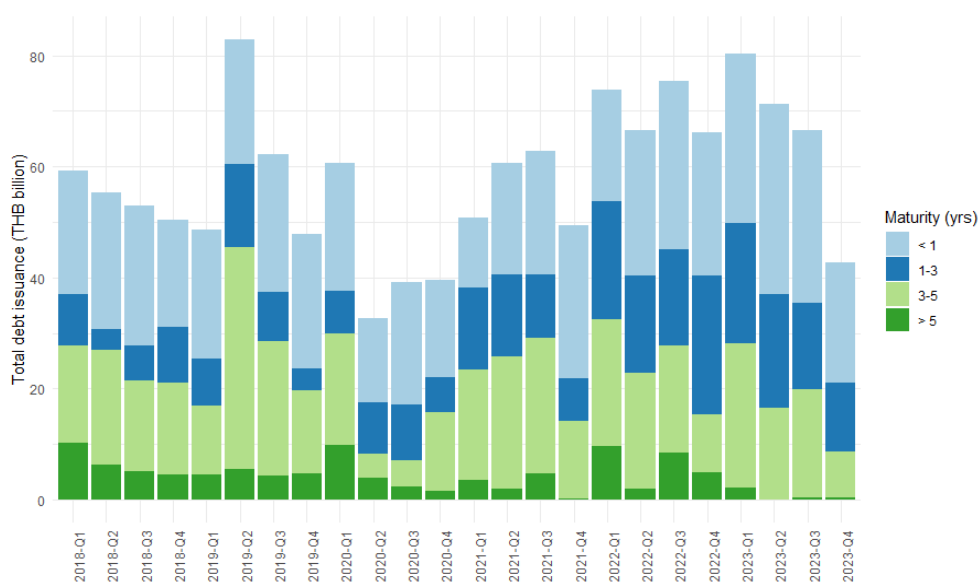

As listed developers acted to avert their crises, they continued to grow. They also continue to issue short-term commercial papers and longer-term bonds, as illustrated in Figure 3. As of 2023Q3, their combined interest-bearing debt is THB 775 million,[10] 54.5% of which is in bonds and commercial papers. About 40% is due in the next 12 months, leaving the sector exposed to rollover risk as before COVID-19 if another funding liquidity draught occurs.

## 5. Conclusion

This paper uses condominium listing data and developers' financial positions to assess the relationship between developers' leverage and fire sales. While observing developers' actions directly is difficult, resale activities are useful to infer their actions. We find evidence consistent with fire sale externalities among listed developers with high leverage but not unlisted developers. Developers' financial fragility seems to be passed on to the secondary market, highlighting the intricate connectivity between the primary and secondary markets for durable assets and the potential for spillover effect. We attribute this difference to the flexibility of bank loan renegotiation versus the rigidity of debt capital market repayments. We also demonstrate that developers' inventory (*excess inventory channel*) and the need for capital market refinancing (*rollover risk channel*) affect their behavior.

---

[9] https://www.thaibma.or.th/Doc/annual/SummaryMarket2020.pdf
[10] The combined number only includes interest-bearing debt reported in company's financial statements. If developers use subsidiaries created as special purpose entities through joint ventures in a way that does not require consolidated financial reporting, their leverage will typically not be included.

When mortgage loans are non-recourse, and property prices are declining, fire sale externalities can increase strategic default, where borrowers who can afford to pay decide to stop paying because it is financially better to be foreclosed and relieved of debt than continue paying, leading to further foreclosures and thus more strategic defaults (Campbell et al. 2011; Gerardi et al., 2015). In Thailand, mortgage loans are effectively full recourse. However, Saengchote and Sampantharak (2022) show that Thai borrowers with multiple types of debt are still more likely to default on mortgage loans if their housing equity is low or negative. While the COVID-19 fire sales externalities did not spill over into the wider financial system, the risk remains as developers continue to use leverage and short-term debt (Figure 3).

In addition, our result points to the success of the Corporate Bond Stabilization Fund (BSF) in averting the rollover crisis, and developers with access to the debt capital market may have even benefited from it as a result. We conclude by highlighting the key roles that commercial banks perform in financial intermediation in the presence of information asymmetry (Townsend, 1979; Campbell and Kracaw, 1980; Diamond, 1984; Fama, 1985) and flexibility in renegotiation during distress (Piskorski, Seru, and Vig, 2010), and the tradeoffs involved in using debt capital market financing for long-duration assets such as real estate. Our findings suggest policymakers can benefit from monitoring developers' leverage and their debt compositions in addition to household leverage for real estate markets.